\documentclass[twocolumn,showpacs,prl,floatfix]{revtex4-1}

\ifx\pdfoutput\undefined
\usepackage{graphicx}
\else
\usepackage[pdftex]{graphicx}
\usepackage{epstopdf}
\fi

\usepackage{subfigure}

\newcommand{\etal}{\emph{et al.}{}}

\begin{document}
\def\Re{\hbox{Re}}
\title{Directed percolation describes lifetime and growth of turbulent puffs and slugs}

\author{Maksim Sipos and Nigel Goldenfeld}
\affiliation{Department of Physics, University of Illinois at Urbana-Champaign, Loomis Laboratory of Physics, 1110 West Green Street, Urbana, Illinois, 61801-3080, USA}

\begin{abstract}

We show that Directed Percolation (DP) simulations in a pipe geometry
in 3+1 dimensions fully capture the observed complex phenomenology of
the transition to turbulence. At low Reynolds numbers (Re), turbulent
puffs  form and spontaneously relaminarize. At high Re, turbulent slugs
expand uniformly into the laminar regions.  In a spatiotemporally
intermittent state between these two regimes of Re, puffs split and
turbulent regions exhibit laminar patches.  DP also captures the main
quantitative features of the transition, with a superexponentially
diverging characteristic lifetime below the transition.  Above the
percolation threshold, active (turbulent) clusters expand into the
inactive (laminar) phase with a well-defined velocity whose scaling
with control parameter (Reynolds number or percolation probability) is
consistent with experimental results.  Our results provide strong
evidence in favor of a conjecture of Pomeau.
\end{abstract}


 \maketitle

The transition from laminar to turbulent flow, studied first by
Reynolds~\cite{reynolds} in pipes, remains a source of complex and
fascinating phenomenology. Varying the dimensionless parameter bearing
his name, $\hbox{Re}\equiv UL/\nu$, where $U$ and $L$ are
characteristic flow velocity and length scales and $\nu$ is the
kinematic viscosity, Reynolds observed localized clusters of
turbulence, now commonly called ``puffs'', that can spontaneously split
or decay. Later, Wygnanski~\etal~\cite{wygnanski1} systematically
described the phase diagram of the laminar to turbulent transition as a
function of the Reynolds number, and the dynamics of the transition is
an active area of investigation
today~\cite{KERS05,eckhardt2007,eckhardt125th}. Laminar pipe flow is
known to be linearly stable for all Reynolds numbers, but small
disturbances can trigger a transition to the turbulent
state~\cite{SALW80}. For sufficiently low $\Re$ the fluid flow is
always laminar and any turbulent disturbances decay immediately.
However, when $\Re \sim 2000$ and the inlet disturbance is large
enough, metastable turbulent puffs can appear. These are regions of
turbulent flow of a fixed size (on the order of pipe diameter $D$)
that can spontaneously relaminarize.

\begin{figure}
  \begin{center}\includegraphics[width=3.5in]{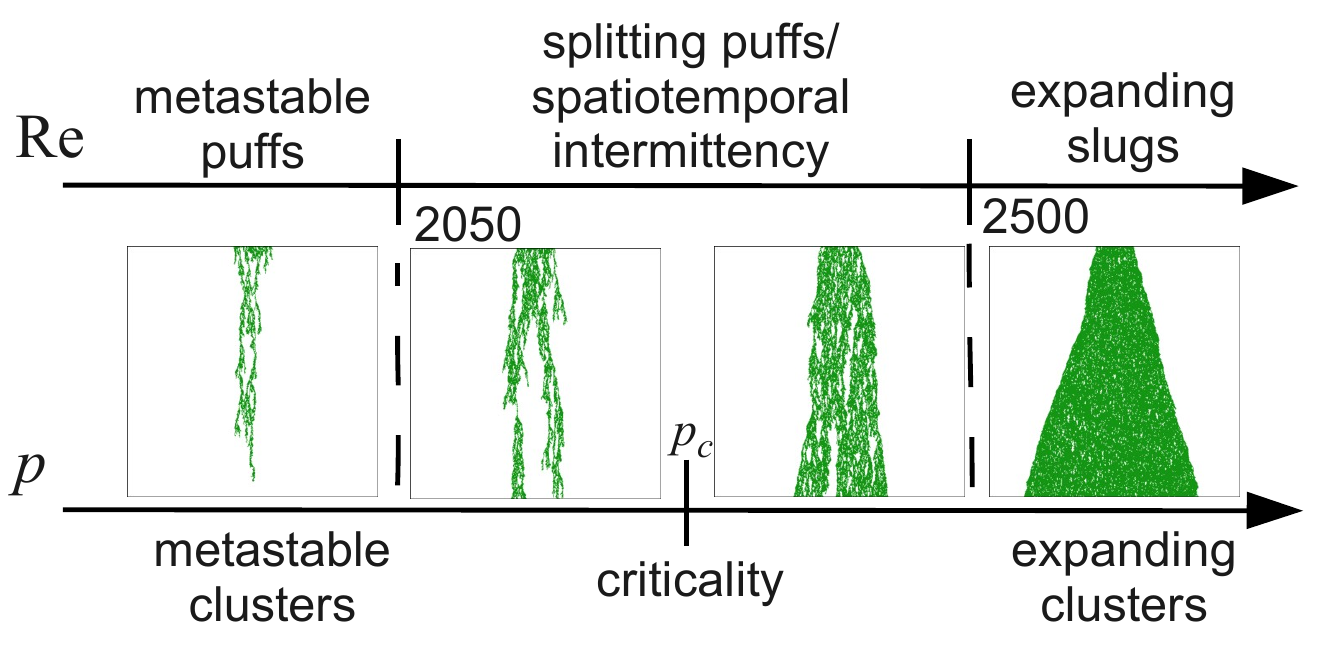}\end{center}
  \vspace{-0.2in}
  \caption{(Color online).  Comparison of the phenomenology of transitional turbulence as a function of Re with that of
  DP in 3+1 dimensions as a function of $p$, both in a pipe geometry.\label{fig:phase}}
\end{figure}

The lifetime of metastable turbulent puffs has been carefully measured
for $1650 < \Re < 2050$, and found to grow superexponentially with Re,
spanning eight orders of magnitude in time~\cite{hof_lifetime}.  For
larger values of $\Re$, the characteristic lifetime of these puffs
grows, and they begin to split and show complex spatiotemporal
behavior~\cite{barkley,hof_splitting}. In this regime, the puffs
interact with a characteristic distance: if puffs come too close to
each other, they may spontaneously
annihilate~\cite{hof_interaction_distance}. The splitting process
continues, until Re exceeds a critical value $\Re_c \sim 2500$, above
which and for a sufficiently large inlet disturbance, a uniform state
of turbulence, a ``slug'', grows with clearly defined turbulent-laminar
interface and a velocity that scales approximately with
$\sqrt{\Re-\Re_c}~$\cite{sreeni, hof_growth_rate}.  The phase diagram
of pipe flow turbulence is shown schematically in Fig.~\ref{fig:phase}.

The purpose of this Letter is to show in detail that the phenomenology
and quantitative details of many features of the laminar-turbulence
transition are consistent with the non-equilibrium phase transition in
the universality class of directed percolation (DP), as originally
conjectured by Pomeau~\cite{pomeau} and continued in subsequent works
by many authors (see, e.g.~\cite{MANN09} and references therein).  Our
work measures the lifetime of active states in DP in a pipe geometry,
finding agreement with the superexponential functional dependence
recently measured by Hof \emph{et al.}~\cite{hof_lifetime}. We also
measure the growth rate of active DP clusters in the supercritical
directed percolation and show that our scaling results are in good
agreement with available experimental data on the growth rate of
turbulent slugs~\cite{sreeni}. These results show that dynamical phase
transition phenomena may be described by directed percolation,
supporting earlier detailed observations of the DP critical exponents
in a fluctuating turbulent liquid crystal system driven by external
forcing~\cite{sano1,sano2}.

The analogy between DP and turbulent-to-laminar transition is the
following: Active states in a three dimensional lattice correspond to
coarse-grained regions of size $\sim \eta$ where the turbulence
intensity exceeds a threshold (here $\eta$ is the viscous scale),
whereas inactive states correspond to patches of the fluid which are
laminar. The dimension along the percolating direction is associated
with time $t$ in the usual interpretation of DP as a dynamical
process. The percolating probability $p$ is analogous to Reynolds
number $\Re$ in the vicinity of the percolation transition, but the
mapping need not be linear.  For the metastable puffs, $\Re < 2050$
region is mapped to $p < p_c$ whereas for the growing fronts, $\Re >
2500$ is mapped to $p > p_c$.  The critical region maps into the
spatiotemporal regime, as summarized in Fig.~\ref{fig:phase}, but this
region and $p_c$ is not strictly defined except in the limit of infinite
system size.  We simulate DP in 3+1 dimensions, where the inlet disturbance
in  pipe flow experiments is modeled as the initial region of active sites.
The simulations are performed in the reference frame of the traveling
puff, which usually travels slower than the laminar mean flow
velocity $U$.

\begin{figure}
  \includegraphics{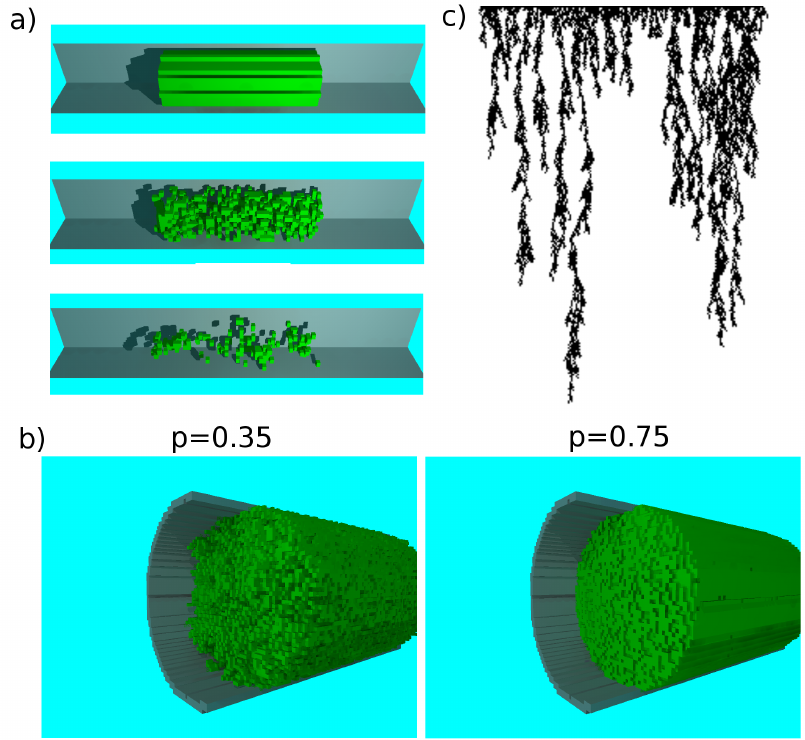}
  \vspace{-0.1in}
  \caption{(Color online) (a) Illustration of the
decay of an active cluster in subcritical directed percolation in 3+1
dimensions in pipe geometry. In this figure the length of the pipe
extends horizontally, and the active sites in percolation are displayed
with green cubes. Inactive sites are not shown. (b) Different front
shapes in the growing front bond DP model. Rough fronts occur for small
$p - p_c$ whereas smoother fronts occur for large $p - p_c$. (c) Time
evolution of an initially active 180 contiguous sites percolating
downwards simulated via bond percolation for $p = 0.61 < p_c$. The
active state (in black) fully decays into the absorbing state (white)
after about 250 steps. \label{fig:3dstuff}}
\end{figure}

\emph{Lifetime of turbulent puffs:} The survival probability of
turbulent puffs in pipe flow is known to be
memoryless~\cite{willis1, eckhardt1},
\begin{equation}
  P(\Re, t) = \exp \left( - \frac{t - t_0}{\tau(\Re)} \right),
  \label{eqn:exponential}
\end{equation}
where $t > t_0$. Here, the survival probability $P(\Re, t)$ refers to
the probability that the turbulent puff still exists after flowing for
time $t$, $t_0$ is the formation time of the puff and $\tau(\Re)$ is the
characteristic lifetime. In Hof \emph{et al}.{}'s work $t_0$ was a
constant (70 $D/U$ where $U$ is the mean flow velocity and $D$ is the
pipe diameter)~\cite{hof_lifetime}. By measuring $P(\Re, t)$ for
specific times $t$ and Reynolds numbers $\Re$, Hof~\etal~calculated
$\tau(\Re)$ from (\ref{eqn:exponential}). They discovered that $\tau$
scales superexponentially with Reynolds number~\cite{hof_lifetime},
fitting well a parameterization of the form:
\begin{equation}
  \tau(\Re) = \tau_0 \exp [ \exp (c_1 \Re + c_2) ],
  \label{eqn:superexponential}
\end{equation}
where $\tau_0 \sim D/U$. 

The survival probability $P(p, t)$ in our DP measurements is the
probability that there will be active sites left in the lattice after
$t$ DP steps. From $P(p, t)$, the lifetime of the disturbance $\tau$
can be measured just as in Hof~\etal. This idea is illustrated in the
snapshots of the simulation in Fig.~\ref{fig:3dstuff}(a). Here, DP is
simulated in 3+1 dimensions in a pipe of radius of 5 lattice sites. In
this simulation $p$ is less than $p_c$, and so the puff eventually
decays. One can measure the lifetime with a 3-dimensional lattice where
two of the spatial dimensions span a disk of radius of $R$ lattice
points (corresponding to the pipe radius), with fixed boundary
conditions. However, the measurement of lifetime in this way over many
orders of magnitude is made difficult because of the system size. The
lifetime measurements must be repeated many times to be able to obtain
sufficient statistics. However, when $p$ is close to the critical
percolation threshold $p_c$, the correlation length $\xi_\perp \sim (p
- p_c)^{\nu_\perp}$ along a space dimension becomes larger than $R$,
and the nominally 3+1 dimensional DP is effectively 1+1
dimensional. Thus, to get sufficient statistics we simulate DP in a
1-dimensional lattice of length $N$ that is initially made to be active
in a subregion of length $N_0$. Below the DP critical point the active
states will eventually decay into the absorbing state. In a
finite-sized system, the decay can always occur, but we find that the
characteristic lifetime $\tau$ of the decay grows super-exponentially
with percolation threshold $p$, and beyond a certain percolation
probability, the average lifetime of the active state is too large to
be measurable on a computer.

\begin{figure}
  \includegraphics[width=0.235\textwidth]{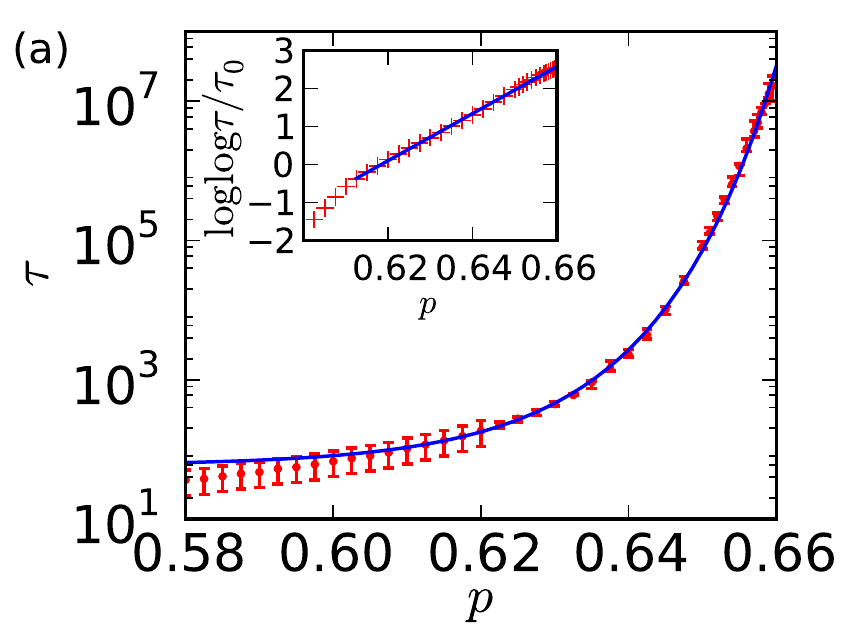}
  \includegraphics[width=0.235\textwidth]{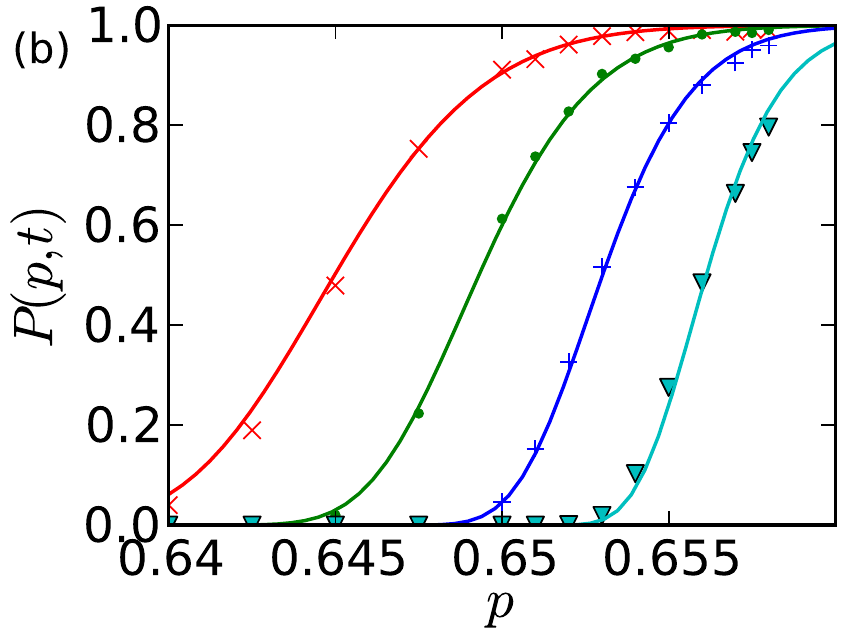}\\
  \includegraphics[width=0.235\textwidth]{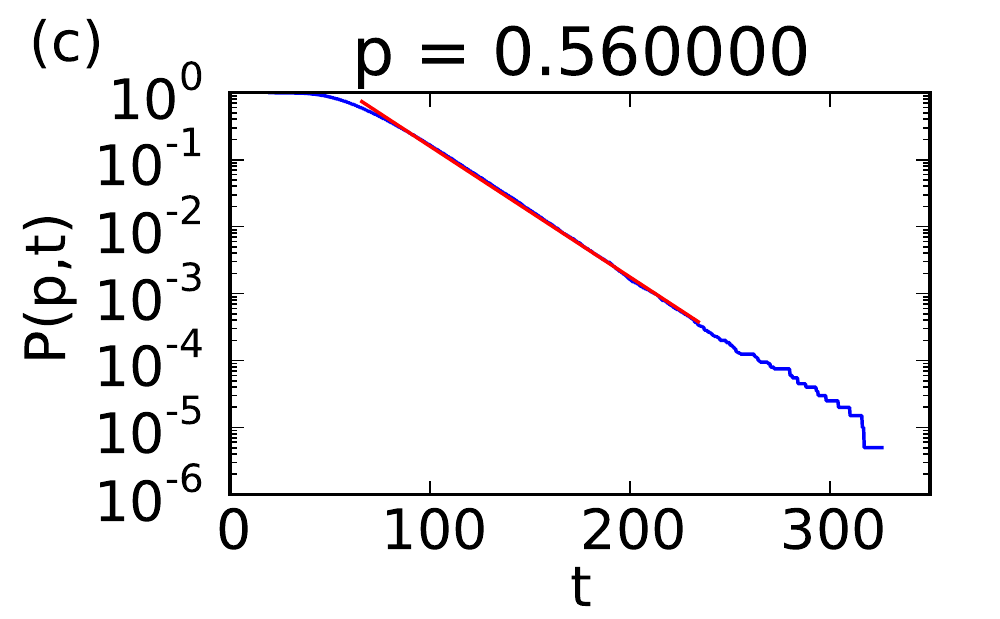}
  \includegraphics[width=0.235\textwidth]{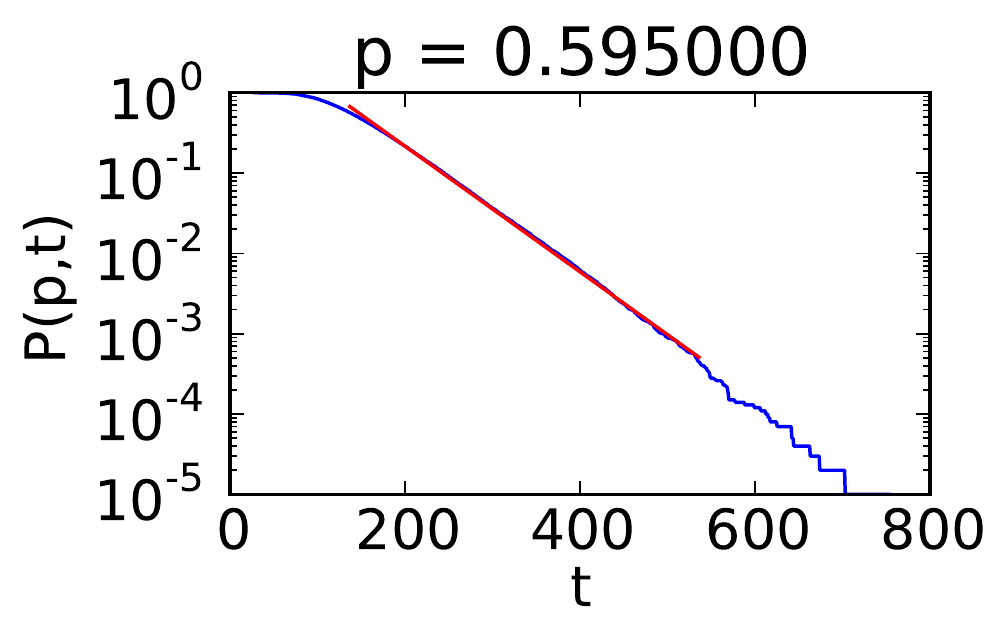}\\
  \includegraphics[width=0.235\textwidth]{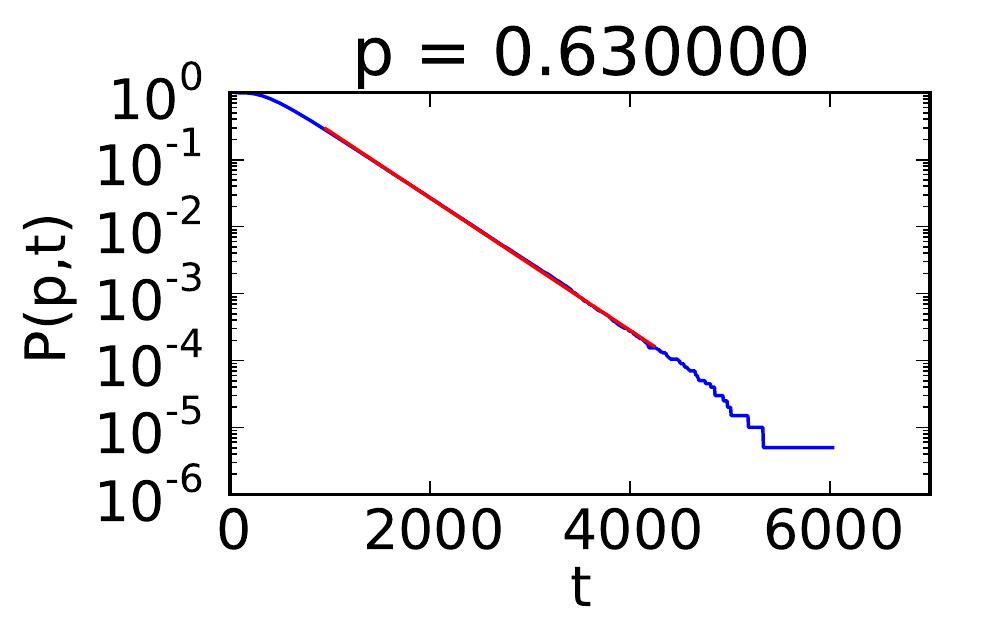}
  \includegraphics[width=0.235\textwidth]{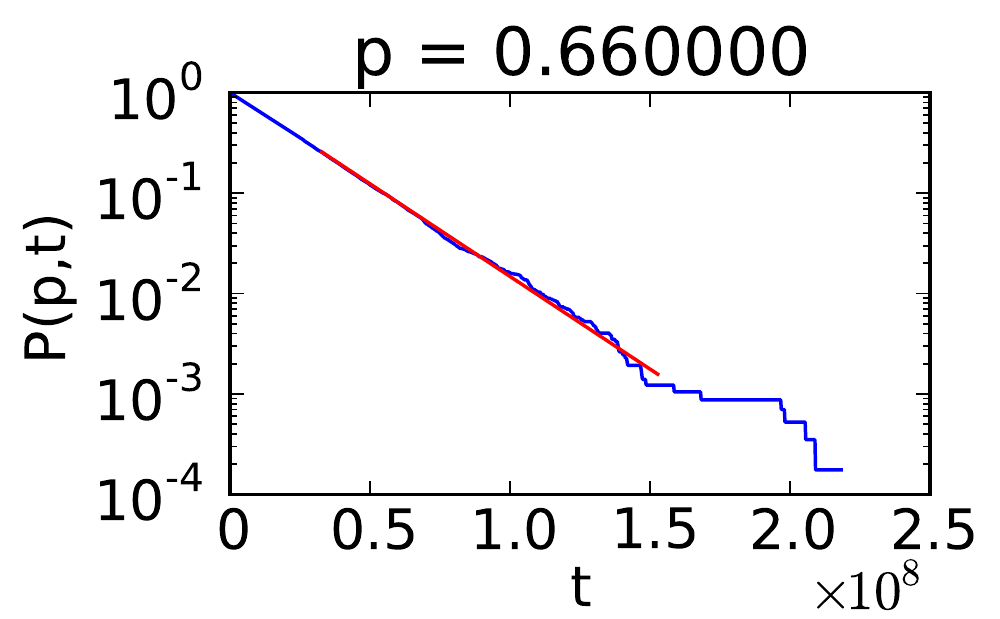}
  \caption{(Color online) (a) Superexponential scaling
  of the characteristic lifetime $\tau$. The line
  indicates the fit to (\ref{eqn:superexponential}).
  Error bars indicate 95\% confidence intervals from
  Kolmogorov-Smirnov test (for $p>0.62$) and 90\%
  confidence intervals from $\chi^2$ test (for $p\leq0.62$).
  In the inset, $\tau_0 = 0.017$.
  (b) Numerical data for survival probabilities $P(p,
  t)$ (points) and a fit to (\ref{eqn:exponential})
  with $\tau$ given by (\ref{eqn:superexponential})
  (solid lines). Data shown for $t=300$ (red crosses),
  $t=1000$ (green dots), $t=6000$ (blue plusses) and
  $t=80000$ (cyan triangles). The value of $p_c$
  observed corresponds to that of 1+1 dimensional bond
  percolation.  (c) Measured survival probabilities as
  functions of $t$ for 4 different values of percolation
  probability $p$.  Blue line indicates measured data,
  whereas red line indicates a fit to the exponential
  distribution.  Deviations from exponential distribution
  for small $t$ are due to nonzero $t_0$. \label{fig:tau}}
\end{figure}

Hof~\etal~were able to calculate $\tau$ via (\ref{eqn:exponential}) by
measuring $P(\Re,t)$. Even though they could only extend $t$ to $3450
D/U$, they were able to resolve $P$ to 100 ppm, giving them effective
measurements of $\tau$ over 8 orders of magnitude. In the case of
directed percolation, one cannot use this procedure, since $t_0$ is not
constant, but instead depends on the percolation probability $p$. In
directed percolation, $t_0$ is the time over which the initial state is
remembered by the system. Hence, for each value $p$ we must measure the
survival lifetimes of many instantiations of directed percolation. The
cumulative distribution function (CDF) of these survival lifetimes then
approximates the survival probability $P(p, t)$, as long as
sufficiently many instantiations have been performed. From the fit of
the form (\ref{eqn:exponential}) to the CDF data, one can read off
$\tau(p)$ and $t_0(p)$. Our measurements of $\tau(p)$ for a lattice of
size $N=100$ and $N_0=20$ are given in Fig.~\ref{fig:tau}(a). The line in
the Figure is obtained by fitting $\tau$ to
(\ref{eqn:superexponential}). The inset shows the linear fit to $\log
\log \tau(p)/\tau_0$. Sufficiently far away from the critical point $p_c$,
we find that the linear fit deviates, indicating that the
superexponential behavior may somehow be related to the diverging
correlation lengths at the critical point.

From the CDF data, we can evaluate the survival probability functions
analogously to Figure 2 of \cite{hof_lifetime}. Figure \ref{fig:tau}(b)
shows our numerical data $P(p, t)$ for 4 different times (S curves),
along with the model $P(p,t) = \exp(-(t-t_0) / \tau(p))$. To evaluate
the model fit, we use the value of $t_0(p_{0.2})$ where $p_{0.2}$ is
found by finding where $P(p_{0.2}, t) = 0.2$. Note that the fact that
the S curves become steeper with $p$ is a characteristic of the
superexponential scaling of $\tau(p)$.

The numerical data presented so far in the paper has been measured in a
finite volume of size $N=100$. Finite size effects in DP have been
investigated thoroughly in the literature~\cite{dp_paper}. We ran our
simulations in a volume bound only by the range of the integers on our
computer (0 to $2^{63}-1$), and we didn't find any qualitative
differences regarding the superexponential scaling of $\tau(p)$.

\emph{Growth rate model:} When $p>p_c$ active DP clusters grow in the
pipe. We measured their growth rate and related it to the growth rate
of turbulent slugs. The speed at which the front of the percolating
clusters propagates into the neighboring inactive region is given by
$G \sim \xi_{\perp} / \xi_{\parallel} \sim (p - p_c)^{\nu_\parallel -
\nu_\perp}$, where $\xi \sim (p - p_c)^{-\nu}$ is the correlation
length in direction of space (denoted by $\perp$) or in direction of
time (denoted by $\parallel$). Using the above prescription and numerical
values of DP critical exponents~\cite{dp_paper} one should expect that
$G \sim (p - p_c)^{\gamma}$ where $\gamma = 0.524$ in 3+1 dimensional
DP,  whereas $\gamma = 0.637$ in 1+1 dimensions. These power laws are
close to the exponent $0.5$ first proposed in 1986~\cite{sreeni}, as well
as in modern experiments~\cite{hof_growth_rate}.  However, the data is
not sufficient yet to differentiate between two such close power law
exponents.

\begin{figure}
  \includegraphics[width=0.235\textwidth]{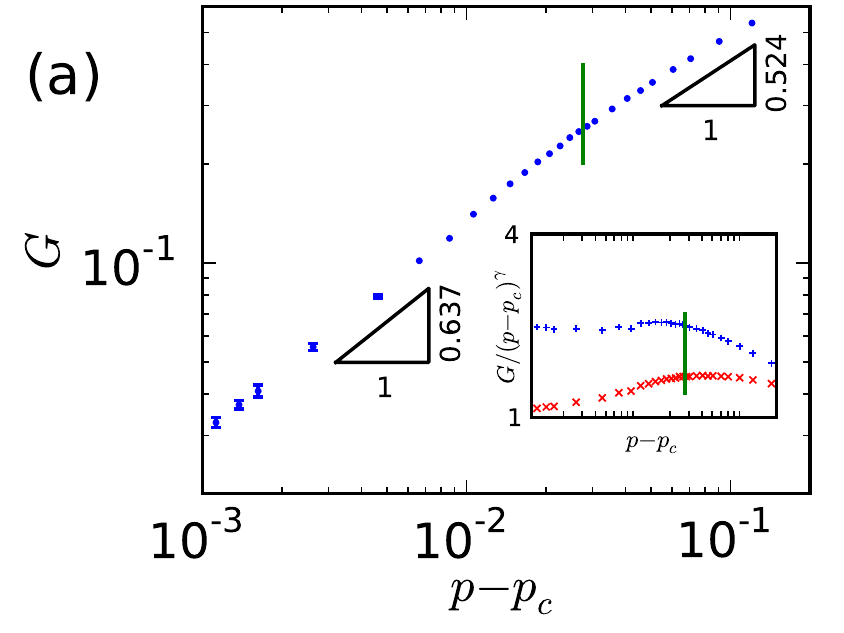}
  \includegraphics[width=0.235\textwidth]{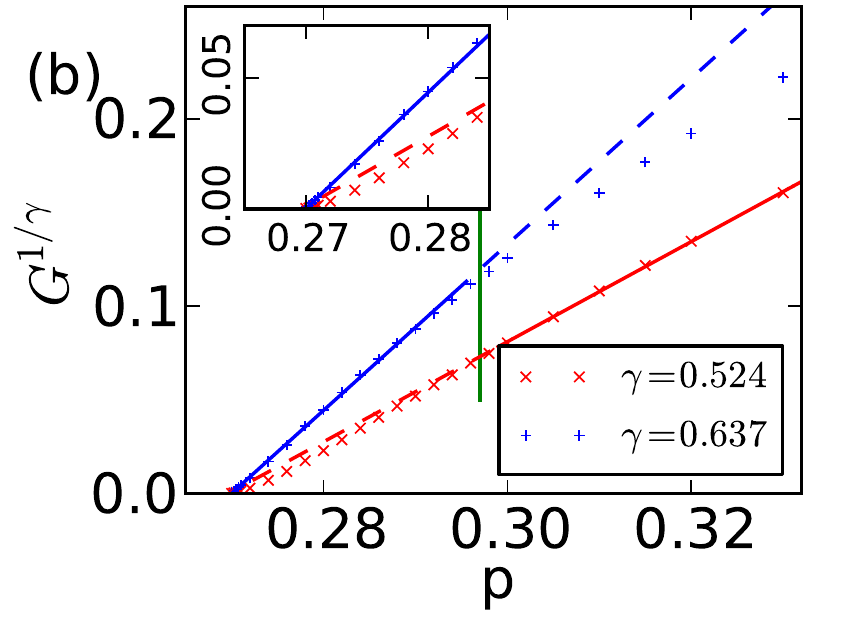}
  \caption{(Color online)
  Measured values of front propagation velocity $G$ (red crosses)
  compared to theoretical prediction for 3+1 dimensional and 1+1
  dimensional DP. Green vertical line indicates the value of $p$ for
  which $\xi_\perp$ exceeds $1.2R$. We call this value $p_R$.
  (a) Plot indicating the power law crossover of $G(p)$. Inset shows
  that $G/(p-p_c)^{0.637}$ is roughly constant for $p<p_R$ and
  similarly $G/(p-p_c)^{0.524}$ for $p>p_R$.
  (b) Both regimes of DP have the same critical point $p_c$ (within the
  error of our measurements).  The linear fits to $G^{1/\gamma}$ are
  shown in solid lines.  Extrapolations are indicated with dashed
  lines, and they cross at the same $p_c$.  This value of $p_c$ is
  corresponds to that of 3+1 dimensional bond percolation.\label{fig:G}}
\end{figure}

Measurements of the growth rate $G$ of an initially active region in
3+1 DP in a pipe geometry are shown in Fig.~\ref{fig:G}. The
measurements were made by simulating bond DP with $p>p_c$ and measuring
the positions of the two fronts as functions of time. During the
numerical simulation we also measure the correlation length $\xi_\perp$
by calculating the root-mean-square height (i.e. roughness) of the
turbulent-laminar interface. The agreement with theoretical expectation
is good, and we see numerical evidence for the crossover from 3+1 to
1+1 dimensions. When $p - p_c \ll 1$, then $\xi_\perp > R$ and the
system is effectively 1+1 dimensional. In Fig.~\ref{fig:G}(b), we have
plotted $G^{1/\gamma}$ versus $p$ for different choices of $\gamma$
corresponding to 1+1 and 3+1 dimensional DP. We see clearly the
crossover between the expected regimes. Note that in this plot we did
not need to guess $p_c$: both scaling regimes yield the same $p_c$. It
is difficult however to extend the data for $G(p)$ close to $p_c$. Due
to the finite size of the system, when $p - p_c$ is small, the active
regions split and may decay into the absorbing state. This makes it
difficult to clearly measure front propagation velocity. On the other
hand, when $p - p_c$ is large, the scaling breaks down.  Thus we expect
the power law exponent of 0.524 to be observable only in an
intermediate regime of $p-p_c$, sufficiently close to $p_c$ but still
such that $\xi < R$.

One other aspect of the phenomenology of pipe flow is captured by the
DP model, namely that the fronts of active regions with $p - p_c \ll 1$
are much rougher than when $p - p_c$ is large. This is because the density
of active states within the region is an increasing function of $p$.
Furthermore, the width of the front is related to the spatial correlation
length $\xi_\perp$ which becomes small when $p-p_c$ is large. The
difference between the rough and smooth front regimes is shown in
Fig.~\ref{fig:3dstuff}(b). This is analogous to the results in pipe
flow experiments hot wire measurements, where puff structures were
found to have rough edges whereas slugs have clearly defined
fronts~\cite{wygnanski1,nishi}.

\emph{Conclusion:} The DP simulations of characteristic lifetime
presented in this paper have been performed via the bond percolation
algorithm. However, we found superexponential scaling of $\tau(p)$ for
site percolation too. Therefore, the superexponential scaling of the
lifetime is likely a universal characteristic of the directed
percolation process. Goldenfeld~\emph{et al.}~proposed that this
superexponential character of the turbulent puff lifetime can be
described by extreme value statistics~\cite{fisher_tippett}, because
puff decay occurs when turbulent energy fails to attain the required
threshold at all points in the puff~\cite{nigel_evs}. In the usual
central limit theorem, under appropriate conditions~\cite{feller} the
distribution of a sum of $N$ random variables tends to a Gaussian
distribution for large $N$. However, a maximum (or minimum) of $N$
random variables is instead superexponentially distributed with 3
universality classes~\cite{fisher_tippett}, selected by the underlying
probability distribution of $\{x_i\}$. In Fig.~\ref{fig:3dstuff}(c)
we show a time evolution of an initially active cluster percolating
with $p < p_c$. The lifetime of the entire cluster is the lifetime of
the longest active ``strand'' percolating downwards. Assuming that
strand lifetimes are independent and identically (exponentially)
distributed, then the lifetime of the longest strand is given by the
type I Fisher-Tippett distribution $\exp(-\exp(-p))$. This argument has
also been used to explain the superexponential distribution of size of
largest connected cluster in ordinary (isotropic)
percolation~\cite{bazant}, and it was found there that correlations
between cluster sizes (analogous to strand lifetimes) do not influence
the superexponential scaling.

The DP model we proposed in this paper can account for the
superexponential lifetime of the turbulent puffs, as well as the
uniform growth rate of turbulent slugs. As shown in
Fig.~\ref{fig:phase}, the transition between these two regimes ($2050 <
\Re < 2500$) occurs through the splitting and interactions of puffs.
The spatio-temporal patterns of coarse-grained turbulent intensity
obtained from a direct numerical simulation~\cite{hof_splitting} bear
similarities to those of directed percolation, but the data are not
adequate to make a quantitative analysis.  Thus, we do not know if
additional hydrodynamic realism is required for a detailed account of
the regime $2040< \Re < 2500$.  After this work was complete, we
learned of a new preprint by Barkley, which uses a coupled map and a
shear field to model the laminar-turbulent transition~\cite{BARK11}.  It
is possible that this model is in the DP universality class also.

We thank M.~Avila and B.~Hof for valuable discussions.  This work was
partially supported by the National Science Foundation under Grant No.
NSF-DMR-1044901.

\bibliographystyle{apsrev}

\bibliography{dpt}

\end{document}